\pdfoutput=1
\documentclass[twocolumn,10pt]{article}

\usepackage[T1]{fontenc}
\usepackage[utf8]{inputenc}
\usepackage{mathptmx}   
\usepackage{microtype}

\usepackage[letterpaper,top=0.85in,bottom=0.95in,left=0.72in,right=0.72in,columnsep=0.28in]{geometry}
\usepackage{titlesec}
\usepackage{titling}
\usepackage{booktabs}
\usepackage{graphicx}
\usepackage{amsmath,amssymb}
\usepackage{caption}
\usepackage{subcaption}
\usepackage{balance}
\usepackage{enumitem}
\usepackage{xcolor}
\usepackage{tikz}
\usetikzlibrary{arrows.meta,positioning,shapes.geometric,calc,fit,backgrounds}
\usepackage[hidelinks]{hyperref}

\definecolor{deepblue}{HTML}{1B4965}
\definecolor{lightblue}{HTML}{8ECAE6}
\definecolor{orange}{HTML}{E07A3F}
\definecolor{green}{HTML}{5AA469}
\definecolor{teal}{HTML}{2A9D8F}
\definecolor{softbg}{HTML}{EEF3F6}
\definecolor{softbg2}{HTML}{FBEEE6}

\titleformat{\section}{\normalfont\large\bfseries}{\thesection}{0.6em}{}
\titleformat{\subsection}{\normalfont\normalsize\bfseries}{\thesubsection}{0.5em}{}
\titlespacing{\section}{0pt}{1.1ex plus 0.4ex}{0.6ex}
\titlespacing{\subsection}{0pt}{0.9ex plus 0.3ex}{0.4ex}

\tikzset{
  idx/.style={rectangle, rounded corners=1.5pt, draw=deepblue, line width=0.7pt,
    fill=softbg, align=center, inner sep=3pt, font=\scriptsize},
  hot/.style={rectangle, rounded corners=1.5pt, draw=orange, line width=0.7pt,
    fill=softbg2, align=center, inner sep=3pt, font=\scriptsize},
  plain/.style={rectangle, rounded corners=1.5pt, draw=black!55, line width=0.6pt,
    fill=white, align=center, inner sep=3pt, font=\scriptsize},
  fl/.style={-{Latex[length=1.6mm]}, line width=0.6pt, draw=black!65},
}

\begin{document}

\twocolumn[{%
\begin{@twocolumnfalse}
\vspace{-1.2em}
\begin{center}
{\LARGE\bfseries CeQe: Grounding Lexical Retrieval\\[2pt]
in Semantic Evidence\par}
\vspace{0.8em}
{\large Adam Kahirov, Umesh Deshpande, Swaminathan Sundararaman\par}
\vspace{0.2em}
{\normalsize\itshape IBM Research\par}
\end{center}
\vspace{0.6em}
\begin{quote}
\noindent\textbf{Abstract.}
Lexical retrieval (BM25) captures exact keyword matches and weights terms by
corpus-wide significance, but it is blind to the semantic vocabulary gap:
when a relevant document phrases an answer differently from the query, BM25
never retrieves it, and no amount of downstream reranking or fusion can
recover a document that was never in the candidate set. We present
\emph{Cross-Encoder Query Expansion} (CE-QE), which reads the per-token
relevance attributions of a cross-encoder applied to top semantic search
results, selects the terms the cross-encoder treats as decisive, and appends
them to the BM25 query. Unlike classical pseudo-relevance feedback, which
reuses BM25's own (possibly wrong) top results, CE-QE seeds expansion from the
semantic retriever's results, avoiding self-reinforcing query drift. Unlike
recent generative query expansion (HyDE, Query2doc), which prompts a large
language model to hallucinate text from its parametric knowledge, every CE-QE
expansion term is copied verbatim from a retrieved passage, so it cannot
introduce vocabulary the corpus does not contain, and its only added cost is
attribution extraction on a cross-encoder a hybrid pipeline already runs for
reranking. On seven BEIR datasets, CE-QE improves lexical recall substantially
where query and answer vocabulary diverge (e.g., NQ Recall@100 from 0.32 to
0.47), and its score-fusion variant (SESF) beats cross-encoder score fusion by
2.5\% on Recall@100 and beats SPLADEv2 and ColBERTv2 by 5.3\% and 4.6\% on
nDCG@10, while leaving the underlying BM25 index completely unmodified.
\vspace{0.3em}

\end{quote}
\vspace{1.0em}
\end{@twocolumnfalse}
}]

\section{Introduction}
\label{sec:intro}
Hybrid retrieval that fuses lexical (BM25) and dense semantic search is now
the default for retrieval-augmented workloads, because each retriever
recovers what the other misses: lexical matching captures exact terms and
corpus-wide term significance, while dense retrieval captures synonyms and
paraphrase \cite{thakur2021beir}. The value of the lexical stage is
concrete---it captures exact matches (a dense model can rank ``Apple'' the
company and ``apple'' the fruit as near-identical) and weights rare,
discriminative terms by their corpus-wide significance---but it has a
structural weakness that fusion alone does not fix.

\textbf{The problem.} BM25 retrieves only documents that share surface terms
with the query. When a relevant document phrases the answer differently
(``side effects'' vs.\ ``adverse reactions'', ``ibuprofen'' vs.\ ``NSAID''),
lexical search misses it entirely, and simply fusing its result list with a
semantic list does not repair the lexical ranking itself: fusion and
reranking can only reorder or combine documents that were \emph{retrieved in
the first place}. If a relevant document shares no surface terms with the
query, BM25 never puts it in the candidate set, and no amount of downstream
re-scoring can recover it---the recall ceiling is fixed at the first stage.

This paper contributes a query-side fix for this problem, together with an
account of why it works and how it compares to the two nearest alternatives.

\begin{itemize}[leftmargin=1.1em,itemsep=1.5pt,topsep=2pt]
\item \textbf{Cross-Encoder Query Expansion (CE-QE)} (Section~\ref{sec:ceqe}):
a query-expansion method that reads the per-token relevance attributions of a
cross-encoder applied to top semantic results, selects the terms the
cross-encoder treats as decisive, and appends them to the BM25 query. This
injects semantic signal into the lexical stage directly, without modifying the
index. Its fusion variant, Semantically Enriched Score Fusion (SESF), combines
the enriched lexical results with semantic scores and delivers the best
overall quality.

\item \textbf{A comparison to the two nearest alternatives}
(Section~\ref{sec:novelty}): classical pseudo-relevance feedback (RM3), which
is self-referential in a way that reinforces exactly the failure CE-QE
targets, and recent generative query expansion (HyDE, Query2doc), which
expands a query with text an LLM hallucinates from its parametric knowledge
rather than from the corpus itself, at the cost of a separate generation call
per query. CE-QE expands the query only with terms already present in
passages the corpus's own semantic retriever returned, using a cross-encoder a
hybrid pipeline already runs for reranking, which grounds every expansion term
in the corpus and adds no separate model call.
\end{itemize}

We evaluate CE-QE and SESF on BEIR through RUMIR, a reproducible pipeline
built for this work (Section~\ref{sec:eval}), against standard flat BM25 as
the lexical stage; a companion paper addresses lexical retrieval's separate
scaling problem at billion-chunk scale, and Section~\ref{sec:discussion}
discusses how the two compose.

\section{Background and Related Work}
\textbf{Lexical retrieval.} BM25 scores a document $D$ for query $Q$ as a sum
over query terms of an IDF weight times a saturated, length-normalized
term-frequency factor \cite{robertson2009prf}. Term saturation (parameter
$k_1$) stops a single repeated term from dominating; length normalization
(parameter $b$) discounts long documents.

\textbf{Dense retrieval and ANN.} Bi-encoders encode text once and search by
similarity \cite{karpukhin2020dpr,reimers2019sbert}, using approximate
nearest-neighbor indexes such as HNSW \cite{malkov2018hnsw} or IVF with
product quantization \cite{jegou2011pq} to remain tractable at scale.

\textbf{Fusion.} Two ranked lists are merged by combining ranks or scores.
Reciprocal Rank Fusion (RRF) sums $1/(k+\mathrm{rank}_i)$ across lists, needs no
tuning, and ignores raw scores \cite{cormack2009rrf}; it is used by
Elasticsearch and LanceDB. Weighted fusion combines normalized scores as
$\alpha\,L(c)+(1-\alpha)\,S(c)$ (OpenSearch, Elasticsearch) but is sensitive to
score normalization because BM25 and cosine live on different scales.

\textbf{Reranking.} Cross-encoders jointly encode the query and each candidate
to produce a precise relevance score; they raise quality but cost far more than
first-stage retrieval, motivating two-stage pipelines that rerank only a small
candidate set \cite{nogueira2019bert}.

\textbf{Query expansion.} Classical pseudo-relevance feedback (e.g., relevance
models / RM3 \cite{lavrenko2001rm}) expands a query with terms that co-occur in
top-ranked documents, using term statistics. CE-QE differs in the selection
signal: it picks expansion terms from a cross-encoder's per-token relevance
attributions rather than from co-occurrence counts, so the added terms are those
a supervised relevance model treats as decisive in the query--passage
interaction. Learned sparse models such as SPLADEv2 \cite{formal2021splade} and
late-interaction models such as ColBERTv2 \cite{santhanam2022colbertv2} attack
the same vocabulary gap by changing the index; CE-QE leaves the BM25 index
unchanged and expands the query instead. A more recent line of work expands
queries \emph{generatively}: HyDE prompts an LLM to hallucinate a hypothetical
answer document and embeds it for dense retrieval \cite{gao2023hyde}, and
Query2doc few-shot-prompts an LLM to write a pseudo-document that is appended
to the query for both sparse and dense retrieval \cite{wang2023query2doc}.
Section~\ref{sec:novelty} contrasts CE-QE with this generative family directly.

\textbf{Evaluation.} BEIR is the standard heterogeneous, zero-shot IR benchmark
\cite{thakur2021beir}. We report Recall@100 (coverage) and nDCG@10 (ranking
quality); the two diverge---two rankings with identical Recall@5 can have very
different nDCG@5 (e.g., 1.0 vs.\ 0.62) depending on where the relevant items
land---so we track both throughout. Retrieval quality of this kind underlies
retrieval-augmented generation broadly \cite{gao2023ragsurvey}, which is the
deployment setting motivating this work.

\subsection{Positioning relative to recent work}
\label{sec:novelty}
\textbf{CE-QE vs.\ generative (LLM) query expansion.} HyDE
\cite{gao2023hyde} and Query2doc \cite{wang2023query2doc} both expand a query
using text an LLM \emph{generates from its own parametric knowledge}: a
hypothetical answer document or pseudo-document that may contain no terms
actually present in the target corpus, and that costs a full LLM decoding pass
per query (typically on the order of a paragraph of generated tokens). CE-QE
expands the query with terms extracted from passages the corpus's own semantic
retriever actually returned, using attribution weights from a cross-encoder
that a hybrid pipeline already runs for reranking (Section~\ref{sec:why-ce}).
This has two concrete consequences. First, grounding: every CE-QE expansion
term is copied verbatim from a retrieved passage, so it cannot introduce
vocabulary the corpus does not contain, whereas a generated hypothetical
document can drift into fluent but corpus-absent phrasing---a documented
failure mode of generative expansion on unfamiliar or ambiguous queries.
Second, cost: CE-QE adds attribution extraction on a component already in the
serving path (Table~\ref{tab:latency} shows the full SESF pipeline, including
this step, at $\sim$557\,ms), whereas generative expansion adds a separate LLM
call whose latency and expense scale with generated length and are incurred
independently of any reranking the pipeline already performs. The two
approaches are not mutually exclusive---an LLM-generated expansion could be
fed through the same attribution-based filtering CE-QE uses to select terms
from it before appending them to the BM25 query---but CE-QE's grounding and
marginal-cost advantages come specifically from sourcing expansion terms from
retrieval rather than generation.

\textbf{CE-QE vs.\ classical pseudo-relevance feedback.} RM3 and relatives
\cite{lavrenko2001rm} expand a query with terms drawn from the top documents
the \emph{same} lexical retriever already returned---self-referential in a
way that hurts exactly the queries CE-QE targets, since if BM25's top results
are already wrong because of a vocabulary gap, feeding terms back from those
same wrong results reinforces the error rather than correcting it (query
drift). CE-QE instead seeds expansion from the \emph{semantic} retriever's top
passages, which remain a trustworthy seed set precisely in the cases where
BM25's own results are not (Section~\ref{sec:ceqe}).

\section{Cross-Encoder Query Expansion}
\label{sec:ceqe}
This section motivates each design choice before describing the mechanism,
because the choices are not arbitrary---each responds to a specific way naive
alternatives fail.

\subsection{Why fix the query, not just the ranking}
\label{sec:why-query}
The standard way to add semantic awareness to a lexical result is to rerank it:
retrieve with BM25, then re-score the candidates with a cross-encoder or fuse
them with a semantic list. Reranking and fusion, however, can only reorder or
combine documents that were \emph{retrieved in the first place}. If a relevant
document shares no surface terms with the query, BM25 never puts it in the
candidate set, and no amount of downstream re-scoring can recover it---the
recall ceiling is fixed at the first stage. This is precisely the vocabulary-gap
failure mode motivating this paper (Section~\ref{sec:intro}): lexical search
retrieves the wrong set, not just ranks the right set poorly. Fixing it
therefore requires intervening \emph{before} retrieval, at the query itself,
so that the missing document has a chance to be retrieved at all. CE-QE is a
query-side fix for exactly this reason: it changes what BM25 searches for,
not how BM25's results are consumed afterward.

\subsection{Why semantic passages, not BM25's own top results, seed expansion}
Classical pseudo-relevance feedback (RM3 and relatives \cite{lavrenko2001rm})
expands the query with terms drawn from the top documents that the \emph{same}
lexical retriever already returned. This is self-referential in a way that
hurts exactly the queries CE-QE targets: if BM25's top results are wrong
because of a vocabulary gap, feeding terms back from those same wrong results
reinforces the error rather than correcting it (query drift). CE-QE instead
seeds expansion from the \emph{semantic} retriever's top passages. Dense
retrieval finds topically related passages by meaning, independent of surface
term overlap, so its top results remain a trustworthy seed set precisely in
the cases where BM25's own results are not. This is the central design
insight: use the retriever that does not have the failure mode to repair the
retriever that does.

\subsection{Why cross-encoder attribution, not raw term statistics}
\label{sec:why-ce}
Given a set of semantically relevant passages, the next question is which of
their words to add to the query. A pseudo-relevance-feedback-style answer would
score candidate terms by frequency or TF-IDF within those passages. This is a
weak signal here: a passage relevant to ``side effects of ibuprofen'' also
contains many generic, high-frequency words (\emph{include}, \emph{common},
\emph{effects} itself) that are not what makes the passage relevant, and
frequency statistics cannot tell the two apart. A cross-encoder, by contrast,
is trained end-to-end to predict query--passage relevance, and its per-token
attributions reveal \emph{which tokens the model actually used} to make that
judgment. For the query ``side effects of ibuprofen'' and the passage ``Common
side effects of ibuprofen include nausea and dizziness'', the cross-encoder
assigns high attribution to \emph{nausea} (0.94) and \emph{dizziness}---content
terms absent from the query---and low attribution to the generic scaffolding
around them. This is a supervised, discriminative-term selector obtained for
free from a component the pipeline needs anyway for reranking, rather than a
separate model that must be trained or tuned.

A second reason to prefer attribution over whole-passage injection is query
compactness. Appending an entire top passage to the query would dilute BM25's
IDF weighting with a burst of common terms and inflate query length past the
point where term-frequency saturation ($k_1$) still discriminates---long,
diffuse queries under-reward the terms that matter. Extracting only a handful
of high-attribution tokens keeps the expanded query short and each added term
individually discriminative, which is what BM25's scoring function is designed
to exploit.

\subsection{The CE-QE algorithm}
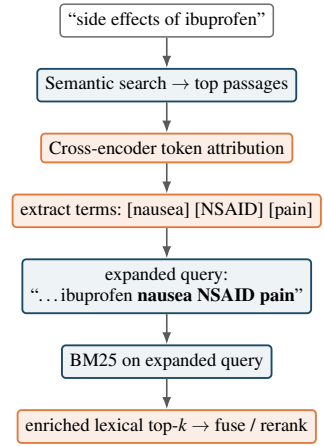
\begin{figure}[t]
\centering
\begin{tikzpicture}[node distance=3.0mm]
\node[plain] (q) {``side effects of ibuprofen''};
\node[idx, below=4mm of q] (sem) {Semantic search $\rightarrow$ top passages};
\node[hot, below=4mm of sem] (ce) {Cross-encoder token attribution};
\node[hot, below=4mm of ce] (tok) {extract terms:\ [nausea] [NSAID] [pain]};
\node[idx, below=4mm of tok] (exp) {expanded query:\\``\dots ibuprofen \textbf{nausea NSAID pain}''};
\node[idx, below=4mm of exp] (bm) {BM25 on expanded query};
\node[hot, below=4mm of bm] (res) {enriched lexical top-$k$ $\rightarrow$ fuse / rerank};
\foreach \a/\b in {q/sem,sem/ce,ce/tok,tok/exp,exp/bm,bm/res}
  \draw[fl] (\a) -- (\b);
\end{tikzpicture}
\caption{CE-QE. Cross-encoder attributions over top semantic passages select
discriminative terms that are appended to the BM25 query, transferring semantic
signal into lexical retrieval without changing the index.}
\label{fig:ceqe}
\end{figure}

CE-QE (Figure~\ref{fig:ceqe}) runs semantic search for the original query,
retrieves the top passages, applies the cross-encoder, and extracts the
highest-attribution content tokens across those passages (e.g., \emph{nausea},
\emph{NSAID}, \emph{pain}). It appends the extracted tokens to the original
query and runs BM25 on the expanded query. In our experiments up to 10 tokens
are extracted from up to 10 documents; both limits are deliberate rather than
incidental. Bounding the number of source passages caps the risk of topic
drift---attribution scores taper quickly past the first few passages, and
terms drawn from lower-ranked, less relevant passages are increasingly likely
to be generic or off-topic---while bounding the token budget keeps the
expanded query short for the reason given in Section~\ref{sec:why-ce}. Terms
are \emph{appended} to the original query rather than replacing it, so the
original terms keep contributing their own IDF weight and BM25 can still
reward an exact match; expansion only adds recall opportunities, it does not
remove the precision the original query already had.

The BM25 index itself is untouched; only the query changes. This is a
deliberate separation of concerns: selecting expansion terms is the one part
of the pipeline that needs a neural model, and it runs once per query over a
small, bounded set of candidate passages (at most 10) rather than over the
corpus, so its cost does not grow with corpus size regardless of which BM25
index answers the expanded query. In this paper we evaluate CE-QE against
standard, flat BM25 (Section~\ref{sec:eval-setup}); we have not run CE-QE
against a hierarchical, billion-scale BM25 index end-to-end. The separation of
concerns is what makes that combination architecturally straightforward---the
expanded query is ordinary BM25 input, so it needs no special handling by any
particular lexical index's internal structure---but we present it as a
compositional argument in Section~\ref{sec:discussion} rather than as a
measured result.

CE-QE targets a specific failure mode. Consider the query ``Brown State Fishing
Lake is in a country that has a population of how many inhabitants?'' whose
answer document is about ``Brown County, Kansas \dots\ the county population was
9{,}984.'' Lexical search fixates on \emph{population}/\emph{country}; semantic
search drifts to \emph{lakes}/\emph{countries}. CE-QE extracts \emph{county},
\emph{kansas}, \emph{census} from the top passages and adds them to the BM25
query, which then matches the answer document that neither stage found alone.

\subsection{Why the benefit concentrates on high vocabulary-divergence datasets}
\label{sec:why-divergence}
CE-QE's mechanism predicts \emph{where} it should help: datasets where queries
and their answer passages are phrased in different registers should benefit
most, and datasets where they already share vocabulary should benefit least.
This matches the evaluation (Section~\ref{sec:eval}). Natural Questions and
TREC-COVID pair informal or lay-phrased queries against encyclopedic or
technical-scientific answer text, respectively---a wide register gap that
expansion closes. FEVER, by contrast, is a fact-verification dataset whose
claims are written to closely paraphrase the Wikipedia sentences they check
against, so query and answer vocabulary already overlap and expansion has
little left to add (and, as Section~\ref{sec:eval} shows, can slightly hurt by
adding terms the original query did not need). This dataset-dependent pattern
is evidence that CE-QE is doing what it is designed to do---closing a
vocabulary gap---rather than improving retrieval through some unrelated,
dataset-independent effect.

\subsection{Semantically Enriched Score Fusion (SESF)}
SESF uses CE-QE as the lexical stage of a score-fusion hybrid: the enriched BM25
results are combined with the semantic results, and the merged set is reranked.
The motivation for fusing rather than using CE-QE's lexical list alone is that
fusion and expansion attack the same problem from different angles and their
gains are not redundant: expansion raises the ceiling on what BM25 \emph{can}
retrieve, while fusion still contributes independent evidence from the
semantic embedding whenever the two retrievers disagree. Because the lexical
list going into fusion is now itself semantically aware, the two lists overlap
more on genuinely relevant documents than plain BM25 $+$ vector fusion would,
giving the fusion step---and the reranker downstream---a stronger and more
concentrated candidate pool to work with. A reranker can only reorder the
candidates it is given, so improving the pool improves the ceiling on what
reranking can achieve, which is why SESF's gains over CE score fusion
(Section~\ref{sec:eval}) show up primarily as higher Recall@100 rather than as
a change in how well already-retrieved items are ranked.

\section{Evaluation}
\label{sec:eval}
\subsection{Setup}
\label{sec:eval-setup}
Experiments run through \textbf{RUMIR}, a pipeline built for this work that
indexes BEIR datasets, executes lexical, semantic, fusion, reranking, and
expansion variants, and evaluates them consistently. Table~\ref{tab:data} lists
the seven BEIR datasets; each configuration is evaluated over 1{,}000 queries at
top-$k{=}100$. Embeddings use \texttt{granite-embedding-30m-english}; the
semantic index is IVF\_PQ (LanceDB) or HNSW (OpenSearch). Rerankers are
\texttt{bge-reranker-v2-m3} (568M params) and \texttt{ms-marco-MiniLM-L12-v2}
(33.4M params, the OpenSearch default). The lexical stage in every result
below---BM25, RRF, CE score fusion, CE-QE, and SESF alike---is standard, flat
BM25 over each dataset's own (at most 8.8M-document) corpus.

\begin{table}[t]
\centering\small
\caption{BEIR datasets used in the evaluation.}
\label{tab:data}
\begin{tabular}{@{}lrl@{}}
\toprule
Dataset & Documents & Domain \\
\midrule
MSMARCO & 8.8M & Web / QA \\
HotpotQA & 5.2M & Wikipedia \\
NQ & 2.7M & Wikipedia \\
FEVER & 5.4M & Wikipedia (fact check) \\
Climate-FEVER & 5.4M & Climate claims \\
TREC-COVID & 171K & Biomedical \\
FiQa & 57K & Finance \\
\bottomrule
\end{tabular}
\end{table}

\subsection{Hybrid fusion vs.\ single retrievers}
\begin{figure}[t]
\centering
\includegraphics[width=0.99\columnwidth]{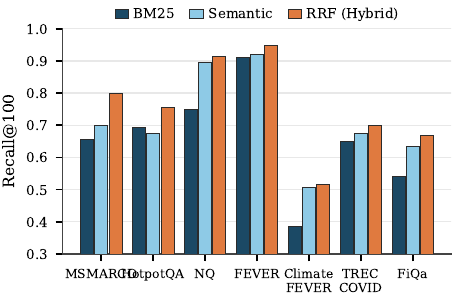}
\caption{Recall@100 for BM25, semantic (IVF\_PQ), and their RRF hybrid across
seven BEIR datasets. Hybrid never loses and often wins substantially, e.g.\
MSMARCO 0.66$\rightarrow$0.80 and NQ 0.75$\rightarrow$0.92.}
\label{fig:fusionrecall}
\end{figure}

\begin{figure}[t]
\centering
\includegraphics[width=0.82\columnwidth]{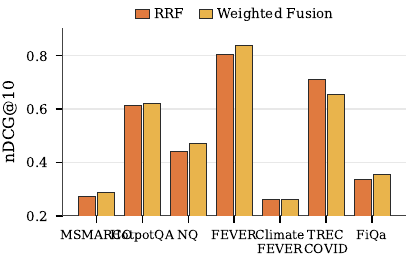}
\caption{RRF vs.\ weighted fusion, nDCG@10. Recall is nearly identical (not
shown); weighted fusion ranks better by retaining score magnitude.}
\label{fig:rrfwf}
\end{figure}

Fusing BM25 with semantic search via RRF improves Recall@100 on every dataset
(Figure~\ref{fig:fusionrecall}): on average $\sim$4.3\% over semantic-only and
$\sim$10\% over BM25-only, with the largest gains where the two retrievers
disagree most (MSMARCO, FiQa). Comparing the two fusion strategies, RRF and
weighted fusion reach near-identical Recall@100, but weighted fusion ranks
better---higher nDCG@10 on five of seven datasets
(Figure~\ref{fig:rrfwf})---because it preserves score magnitude that RRF
discards.

\subsection{Two-stage reranking: gain and cost}
\begin{figure}[t]
\centering
\includegraphics[width=0.82\columnwidth]{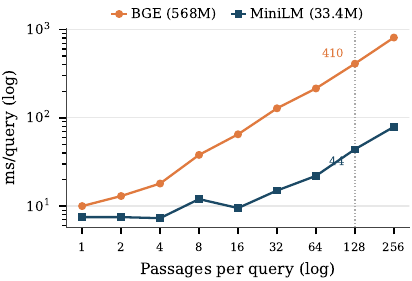}
\caption{Cross-encoder reranking latency vs.\ passages per query (A100, log--log).
BGE is an order of magnitude slower than MiniLM at every candidate-set size.}
\label{fig:celat}
\end{figure}

Adding a cross-encoder reranker (CE score fusion) over the fused candidates
raises quality, and the gain is larger in nDCG@10 than in Recall@100---the
reranker mostly reorders items already retrieved rather than surfacing new ones.
On TREC-COVID, for instance, nDCG@10 rises from 0.71 (RRF) to 0.79 while
Recall@100 rises from 0.70 to 0.83. The cost is the cross-encoder pass, which
scales with the number of passages reranked (Figure~\ref{fig:celat}): at 128
passages, MiniLM costs 43.5\,ms per query but BGE costs 409.6\,ms on an A100---a
$\sim$9$\times$ gap that makes reranker choice the dominant latency lever once
the candidate set is fixed.

\begin{figure}[t]
\centering
\includegraphics[width=0.82\columnwidth]{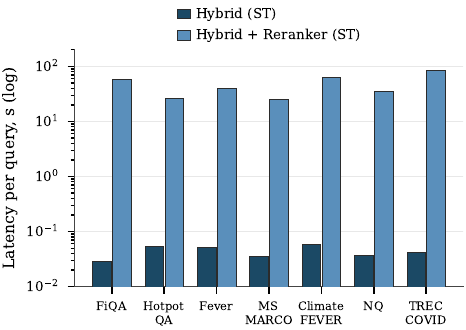}
\caption{Per-query latency, hybrid retrieval alone vs.\ hybrid retrieval with
full-candidate-set cross-encoder reranking (single-threaded, log scale, 7 BEIR
datasets). Unbounded reranking costs 485--2051$\times$ more than retrieval
alone, motivating the bounded candidate sets used elsewhere in this paper.}
\label{fig:rerankoverhead}
\end{figure}

That per-passage cost compounds quickly if reranking is applied naively over a
full, unbounded candidate set rather than a small one. We measure this
directly, single-threaded, with the MiniLM-12L reranker (118M parameters) over
the full hybrid candidate list for each of the seven BEIR datasets, comparing
hybrid retrieval alone against hybrid retrieval followed by full-candidate-set
reranking (Figure~\ref{fig:rerankoverhead}). Hybrid retrieval alone answers in
29--58\,ms per query across datasets; adding full reranking raises this to
25--86\,s per query---a 485--2051$\times$ slowdown, worst on TREC-COVID (86.17\,s,
$2051\times$) and best on HotpotQA (26.21\,s, $485\times$), tracking each
dataset's average candidate-set size more than any other factor. This is the
result that motivates confining reranking to a small, bounded candidate set
rather than the full retrieved list: SESF and CE score fusion both rerank only
the fused top-$k$ (Table~\ref{tab:latency}), which is why their end-to-end
latency lands in the hundreds of milliseconds rather than tens of seconds.

\subsection{CE-QE: enriching the lexical stage}
\begin{figure}[t]
\centering
\includegraphics[width=0.82\columnwidth]{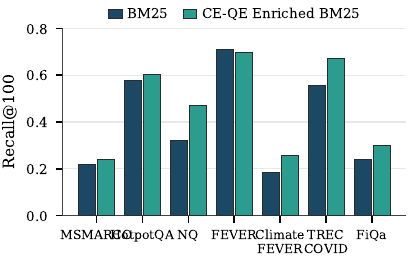}
\caption{CE-QE enriched BM25 vs.\ plain BM25, Recall@100. Expansion terms drawn
from cross-encoder attributions lift recall most where surface vocabulary and
answer vocabulary differ (NQ, TREC-COVID, Climate-FEVER).}
\label{fig:qelex}
\end{figure}

CE-QE improves BM25 directly (Figure~\ref{fig:qelex}). The largest gains appear
where lexical and semantic retrieval otherwise diverge: NQ Recall@100 rises from
0.32 to 0.47, TREC-COVID from 0.56 to 0.67, and Climate-FEVER from 0.19 to 0.26.
FEVER dips slightly (0.71$\rightarrow$0.70) where surface terms already match, so
added tokens contribute little. Because CE-QE only rewrites the query, these
gains require no change to the underlying BM25 index---here, standard flat
BM25---and the same argument applies unchanged if a different, larger-scale
BM25 index answers the expanded query instead (Section~\ref{sec:discussion}).

\subsection{SESF: fusion quality and comparison to SOTA}
\begin{figure*}[t]
\centering
\includegraphics[width=0.80\textwidth]{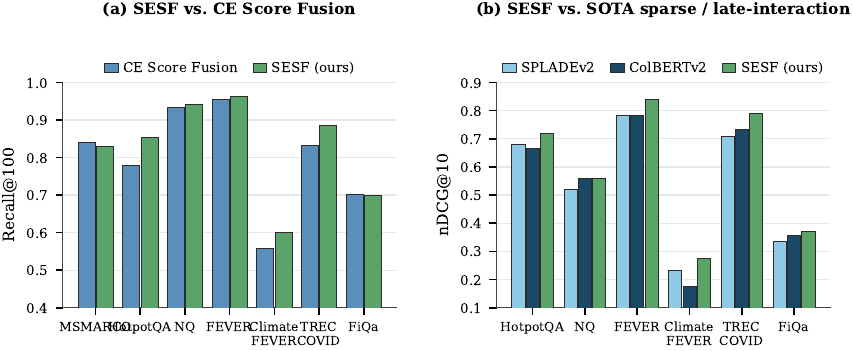}
\caption{SESF quality. (a) SESF vs.\ cross-encoder score fusion, Recall@100:
$+2.5\%$ on average. (b) SESF vs.\ SPLADEv2 and ColBERTv2, nDCG@10: $+5.3\%$ and
$+4.6\%$ respectively, using an unmodified BM25 index.}
\label{fig:sesf}
\end{figure*}

Using CE-QE inside score fusion (SESF) beats cross-encoder score fusion by 2.5\%
on Recall@100 at comparable nDCG@10 (Figure~\ref{fig:sesf}a), with the clearest
wins on HotpotQA, TREC-COVID, and Climate-FEVER. Against strong learned models,
SESF beats ColBERTv2 by 4.6\% and SPLADEv2 by 5.3\% on nDCG@10
(Figure~\ref{fig:sesf}b), while leaving the lexical index a standard BM25 index
rather than a specialized sparse or late-interaction structure. Overall, query
expansion with cross-encoders improves 6.6\% over RRF, the common production
default.

\subsection{End-to-end latency}
Table~\ref{tab:latency} places the query-time cost of each method
(top-$k{=}100$, MSMARCO). RRF is cheapest because it fuses ranks only; CE score
fusion adds the reranker pass; SESF adds semantic retrieval and cross-encoder
token extraction on top, trading $\sim$400\,ms of extra latency for its recall
and SOTA-beating ranking. The extra cost is spent on CPU-friendly BM25 and a
bounded reranker call, not on scaling the index.

\begin{table}[t]
\centering\small
\caption{End-to-end query latency (top-$k{=}100$, MSMARCO) and the quality each
method buys.}
\label{tab:latency}
\begin{tabular}{@{}lcl@{}}
\toprule
Method & Latency & Quality note \\
\midrule
RRF & $\sim$153.8\,ms & fusion baseline \\
CE score fusion & $\sim$400.3\,ms & $+$reranker; better nDCG@10 \\
SESF (ours) & $\sim$556.9\,ms & best recall; beats SOTA nDCG@10 \\
\midrule
\multicolumn{3}{@{}l@{}}{\footnotesize Reranker cost @128 passages: MiniLM 43.5\,ms, BGE 409.6\,ms (A100).}\\
\bottomrule
\end{tabular}
\end{table}

\section{Discussion and Limitations}
\label{sec:discussion}
\textbf{Composing with a scalable lexical index.} CE-QE is evaluated here
against standard flat BM25 over BEIR-scale corpora (at most 8.8M documents);
we have not run it against a billion-chunk lexical index end-to-end. What
makes that composition architecturally straightforward is exactly the
separation of concerns in Section~\ref{sec:ceqe}: CE-QE's output is an
ordinary, expanded text query, not a modified index, a custom scoring
function, or a change to term statistics, so any BM25 index---including a
sharded or hierarchical one---answers it without special handling. We would
expect CE-QE's recall gain to hold at scale, since it is a property of the
query, not of how the index is internally partitioned, but we present this as
an architectural expectation, not a measured result, and confirming it
end-to-end is the natural next step.

\textbf{Cost is reserved for queries that need it.} CE-QE adds a
semantic-search plus cross-encoder pass at query time
(Table~\ref{tab:latency}), so it trades latency for quality and is best
applied selectively rather than unconditionally---on datasets whose surface
terms already match the answer vocabulary (FEVER), the expansion adds cost for
little gain (Section~\ref{sec:why-divergence}). A deployment could use a
cheap signal (e.g., lexical-semantic agreement on the unexpanded query) to
decide when expansion is worth its cost, though we have not evaluated such a
gate here.

\textbf{Attribution quality bounds expansion quality.} CE-QE's expansion terms
are only as good as the cross-encoder's attributions; a reranker with poorly
calibrated or noisy token-level attributions would select less discriminative
terms, and we have not studied sensitivity to reranker choice beyond the two
models used in Section~\ref{sec:eval}.

\section{Conclusion}
Lexical retrieval's blindness to the semantic vocabulary gap cannot be fixed
by reranking or fusion alone, because both operate only on documents already
retrieved. CE-QE fixes it at the source by importing a cross-encoder's
token-level relevance judgments into the BM25 query, grounding every expansion
term in a passage the corpus's own semantic retriever actually returned rather
than in text a language model hallucinates, and adding no separate generation
call. Lexical recall improves substantially where query and answer vocabulary
diverge (e.g., NQ Recall@100 from 0.32 to 0.47), and its fusion form, SESF,
beats cross-encoder score fusion by 2.5\% on Recall@100 and beats SPLADEv2 and
ColBERTv2 by 5.3\% and 4.6\% on nDCG@10, all while leaving the underlying BM25
index completely unmodified---a property that should let it compose with any
lexical index, including one built for scale, without architectural change,
pending direct measurement of that composition.


\balance
\end{document}